# A Blockchain-Based Framework for Distributed Agile Software Testing Life Cycle

MUHAMMAD SHOAIB FAROOQ[1], FATIMA AHMED[2]
[1]Department of Computer Science, University of Management and Technology, Lahore 54000, Pakistan
[2]Department of Computer Science, Bahria University Lahore Campus, Lahore 54000, Pakistan
Corresponding author: Muhammad Shoaib Farooq (shoaib.farooq@umt.edu.pk)

**ABSTRACT** A blockchain-based framework for distributed agile software testing life cycle is an innovative approach that uses blockchain technology to optimize the software testing process. Previously, various methods were employed to address communication and collaboration challenges in software testing, but they were deficient in aspects such as trust, traceability, and security. Additionally, a significant cause of project failure was the non-completion of unit testing by developers, leading to delayed testing. This paper integration of blockchain technology in software testing resolves critical concerns related to transparency, trust, coordination, and communication. We have proposed a blockchain based framework named as TestingPlus. TestingPlus framework utilizes blockchain technology to provide a secure and transparent platform for acceptance testing and payment verification. By leveraging smart contracts on a private Ethereum blockchain, TestingPlus can help to ensure that both the testing team and the development team are working towards a common goal and are compensated fairly for their contributions.

## I. INTRODUCTION

As the base technology for cryptocurrencies like Bitcoin, blockchain technology has come to be recognised for its potential to revolutionise a number of industries outside of banking. A blockchain, at its heart, is a decentralised, distributed ledger that securely, transparently, and immutably logs transactions [1]. A chain of blocks is created by connecting every transaction, or "block", cryptographically to the one before it. Without the use of middlemen or centralised agencies, its decentralised design ensures participant confidence and transparency.

Software testing entails a thorough analysis of a software application or system to detect flaws, problems, or vulnerabilities. Prior works claim that it includes a number of tasks such as development of test case, execution of tests, defect tracking, and reporting. The performance, stability, and safety of the software can be validated through extensive testing, giving quality assurance teams and software engineers the certainty that it satisfies user expectations.

Even though software testing plays a crucial role, traditional techniques frequently run into a number of issues that reduce the efficacy and effectiveness of the testing process. These difficulties include a lack of trust, problems with coordination, gaps in communication, and security worries. Due to the role of numerous stakeholders, including developers, managers of projects, and clients, communication gaps can develop, resulting in misconceptions and lags in addressing crucial issues. When various teams operate in isolation from one another, coordination problems can arise, leading to ineffective test planning, insufficient coverage, and duplication of effort [2]. For organisations and end users, the shortcomings of traditional software testing approaches can have serious ramifications. Unfound flaws and vulnerabilities can have expensive repercussions, including software errors, system failures, data breaches, and monetary losses. Inadequate testing can also lead to bad user experiences, irritation, and a decline in confidence in the product or the company that created it. If word of negative experiences and inferior products spreads via user evaluations and word-of-mouth, businesses' reputations may suffer. As the base technology for cryptocurrencies like Bitcoin, blockchain technology has come to be recognised for its potential to revolutionise a number of industries outside of banking. A blockchain, at its heart, is a decentralised, distributed ledger that securely, transparently, and immutably logs transactions. A chain of blocks is created by connecting every transaction, or "block," cryptographically to the one before it. Without the use of middlemen or centralised agencies, its decentralised design ensures participant confidence and transparency.

Traditional software testing procedures have difficulties, which can be solved by the special advantages provided by blockchain technology. A fool-proof and fair record of all testing actions, including test plans, test cases, test results, and communication, is first provided by blockchain. Blockchain's immutability guarantees that information recorded once cannot be changed or removed, giving a trustworthy and auditable record of testing procedures. Transparency may increase stakeholder trust and facilitate traceability, making it simpler to identify the root causes of flaws and better upcoming testing procedures. Whilst blockchain technology initially sprang to attention as a result of its connection to cryptocurrencies, it has since developed to provide safe and decentralised solutions in a number of different industries. Supply chain management, healthcare, banking, and logistics are just a few of the sectors looking into how blockchain technology could improve their operations' efficiency, security, and transparency. Similar to this, incorporating blockchain technology into software testing offers a chance to get beyond the drawbacks of conventional approaches and streamline the testing procedure.

The TestingPlus framework is a cutting-edge strategy that uses blockchain technology to streamline the life cycle of software testing. It tries to overcome major issues with traditional testing approaches' fundamental lack of coordination, transparency, and trust [3]. TestingPlus integrates blockchain technology to offer an open and transparent platform for payment verification and acceptance testing, fostering productive communication between development teams, testing teams, and other players.

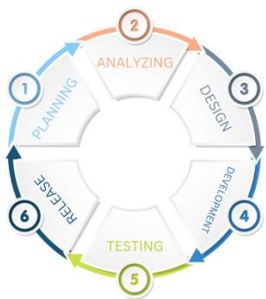

The main goals of the TestingPlus framework are to improve software testing processes' effectiveness and efficiency, collaboration and communication among stakeholders, fair compensation for testing efforts, and the provision of a reliable mechanism for validation and tracking. The use of blockchain technology for safe and fair recording of testing operations, smart contracts for automated and just compensation, and real-time testing progress tracking are some of TestingPlus' core features. Blockchain technology integration throughout all phases of the software testing life cycle is covered by TestingPlus. Blockchain makes ensuring that all testing activities are documented, timestamped, and kept in a secure and decentralised manner, from test planning to development of test case, execution, and tracking of defects. Using smart contracts makes automated payment processes easier and ensures that the testing team and other parties involved are fairly compensated.

To sum up, the use of blockchain technology in software testing is a huge step forward in tackling pressing issues and streamlining the testing procedure. Traditional approaches have had trouble with coordination problems, communication problems, and trust concerns, which has resulted in less-than-ideal testing results. However, the TestingPlus framework, which was put forth in this study, makes use of blockchain technology's advantages to improve software testing's transparency, coordination, and communication. TestingPlus assures fair compensation and fosters effective collaboration between testing teams, development teams, and other stakeholders by offering a safe and fair network for acceptance testing and payment verification. In order to improve software testing procedures and the general quality and dependability of software applications, this research study covers the main features, advantages, and execution considerations of TestingPlus.

## II. RELATED WORK

The quality, reliability, and functionality of software applications depend upon the testing phase of the software development life cycle. In terms of software testing, the emergence of blockchain technology has presented both new challenges and new opportunities. The purpose of this literature review is to offer a thorough examination of the available studies on software testing and their relationship to blockchain technology. This review analyses and synthesizes the results of many research to highlight the pros and cons and limits of various methodologies and to establish the necessity of a comprehensive structure such as TestingPlus.

The requirement for more rigorous quality assurance techniques has contributed to considerable changes in the software testing industry throughout the years. Research on software testing from 2000-2014 was examined in depth by Orso and Rothermel [2]. Automated testing, test case generation, and test execution were only some of the methods and processes they investigated. They brought attention to the shift from labour-intensive manual testing to more time- and labour-saving automated methods. However, the evaluation does not go into detail about the difficulties of software testing in blockchain-based systems or the integration of blockchain technology. The software testing industry faces new hurdles with the introduction of blockchain technology. Several difficulties and solutions for testing blockchain-centric software were found by Koul [7]. Testing the integrity and immutability of blockchain data, validating smart contracts, ensuring interoperability in multi-chain systems, and dealing with scalability issues are all examples of the difficulties that must be overcome. Koul's research sheds light on these difficulties and suggests ways forward. But additional research and specialized frameworks are needed to address these difficulties directly in the context of software testing. Potential benefits of incorporating blockchain technology into software testing include enhanced traceability, security, and transparency. Yau and Potdar [8] developed a blockchain-based testing strategy for team-based software engineering. The team's work stressed the need of teamwork in testing, using the distributed ledger technology of blockchain to improve collaboration and communication between testing groups. They emphasized the benefits of blockchain technology in minimizing reliance on centralized testing infrastructure and facilitating secure and transparent exchange of test results. However, the requirements and difficulties of software testing in blockchain-based environments are not thoroughly investigated in their work. The creation of blockchain-based applications requires thorough testing and analysis of smart contracts. Sujeetha and Perumal [4] researched on testing and analysis of smart contracts. Issues including smart contract efficiency, accuracy, and security were investigated. They stressed the need for thorough testing procedures and instruments to guarantee the safety and dependability of smart contracts. While their research helps fill in some of the gaps in our knowledge of smart contract testing, it does not suggest a testing framework for the full scope of software testing in blockchain-based systems. In the context of blockchain technology, certain drawbacks and limitations of earlier approaches to software testing have been recognized. Using blockchain technology and SDN architecture, Islam et al. [1] presented the SmartBlock-SDN framework for IoT resource management. There is a lack of attention to software testing procedures in blockchain-based applications, despite the fact that their activity is essential to IoT resource management. Their framework focuses mostly on managing and allocating resources without taking into account the difficulties and prerequisites of software testing.

The potential application scenarios and benefits of blockchain technology were highlighted in a recent analysis of blockchain applications by Ghosh et al. [9]. While their work highlights blockchain's potential applications in healthcare, it doesn't provide a testing framework designed for blockchain-based healthcare systems or address the difficulties inherent in software testing. An important area that needs more investigation within the context of software testing is the assessment of the accuracy, reliability, and security of healthcare data recorded on the blockchain. The limitations of SDLC models for blockchain-enabled smart contract applications were explored by Miraz and Alam [6]. The SDLC models for creating blockchain-enabled smart contract applications were the primary focus of their research.

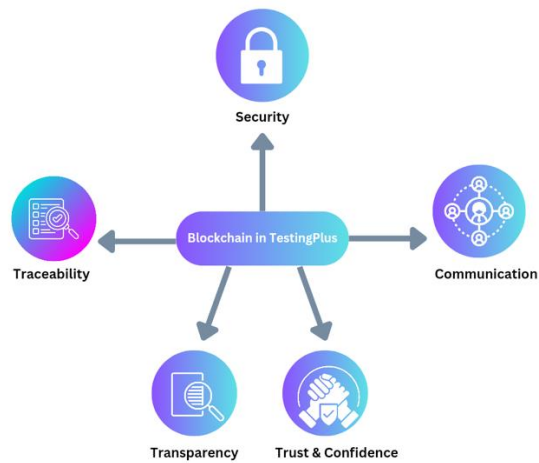

**Fig 2. Blockchain Features in TestingPlus**

When applied to blockchain-based projects, they revealed flaws in the current SDLC models, such as a lack of particular guidance on testing smart contracts and the requirement for increased security. While their work does address difficulties in software development, it does not go further into the complexities of software testing in blockchain settings. The application of blockchain technology to life cycle evaluation is the primary emphasis of Zhang et al [4]. To improve life cycle assessment processes' transparency and traceability, they suggest a blockchain-based implementation methodology and system architecture. However, the difficulties and needs of software testing in the blockchain-based life cycle analysis are not directly addressed in the research. As a result, studies focusing specifically on software testing and future testing frameworks in this space are needed. The research conducted by Kulkarni [5] demonstrates how blockchain technology can be used to track security needs in massive, complicated software development projects. Their strategy concentrates on implementing blockchain technology to safeguard security needs all through the SDLC. However, the study did not go into the unique difficulties and factors that must be taken into account while testing software in this setting. Security testing can be improved and compliance with security criteria can be validated with the help of blockchain technology, but further study is needed to determine how best to integrate blockchain into testing processes.

Given the limitations of existing methods, a comprehensive framework that addresses the particular challenges and conditions of software testing for blockchain applications is desperately needed. We provide the TestingPlus framework as a solution to this problem. Concerns like communication, collaboration, trust, and security are at the heart of TestingPlus' mission to better integrate the benefits of blockchain technology with software testing procedures. TestingPlus provides a secure and open environment for acceptance testing and payment verification by taking advantage of the immutability, transparency, and decentralization of the blockchain. By using smart contracts to automate the execution of established rules and agreements, the framework guarantees that testers will be fairly compensated and builds confidence between the testing and development teams. Enhanced transparency and traceability are two of TestingPlus' most valuable features. All processes for testing and the results are recorded on a blockchain that can be audited but cannot be altered. This transparency inspires confidence among the participants and makes it easy to validate both the testing procedures and the outcomes. In addition to this, TestingPlus makes it easier for testing and development teams to work together effectively. The process of allocating work for testing is streamlined through the use of smart contracts, which also guarantee that all parties involved are reimbursed fairly and in accordance with the terms that have been agreed upon. Because of the automation, the testing procedure has been streamlined, and there have been far fewer instances of arguments or misunderstandings amongst the people involved. In addition, TestingPlus enhances test safety by taking advantage of the built-in protections provided by blockchain technology. Due to the immutable and secure nature of the blockchain, no one will be able to change or modify the results of any tests. As a result of this enhancement in the testing process's integrity and reliability, there is an increased level of trust in the overall quality of the software that has been put through its tests. Additionally, TestingPlus can be used in a variety of commercial applications. It can be used, for instance, to assess blockchain-based healthcare systems, enhancing data security and accuracy and promoting improved communication between healthcare professionals [11]. TestingPlus has another use in supply chain management, where it may confirm the validity of transactions stored on the blockchain to increase system transparency and confidence.

This review of the literature has discussed the subject of software testing and how it relates to blockchain technology. The results have emphasized the positives, negatives, and limitations of current methodologies, highlighting the necessity of a complete framework like TestingPlus. TestingPlus optimizes software testing by integrating blockchain technology to increase transparency, efficiency, and security. By addressing the unique difficulties involved in software testing in blockchain-based systems, it presents a promising approach for boosting the quality and reliability of software applications. TestingPlus has a lot to gain from future advances as blockchain technology continues to expand, resulting in significant innovations in software testing procedures.

## IV. THE PROPOSED FRAMEWORK: TESTINGPLUS

Maintaining the integrity of software is increasingly important in today's dynamic digital environment. Trust, transparency, collaboration, and communication issues plague conventional approaches to software testing, often resulting in ineffectiveness and inferior software quality. By incorporating blockchain technology into software testing, TestingPlus presents itself as a viable option. TestingPlus solves these problems and creates a trustworthy, collaborative, and accountable environment for software testing by exploiting the decentralised, immutable, and transparent properties of blockchain. The importance of blockchain for improving the reliability and auditability of software testing procedures is highlighted by previous researches [8]. By using blockchain, testing processes, test cases, and results are recorded and maintained in an immutable and tamper-proof way, lowering the potential for data tampering and ensuring the legitimacy of testing outputs. In addition, smart contracts are built into TestingPlus to ensure that agreements made between the testing team and the development team are carried out and upheld automatically. The elimination of potential conflicts and misunderstandings is another benefit of this function, which streamlines the coordination of testing activities. Other works show that blockchain technology has the potential to improve software testing by increasing transparency and trust. TestingPlus improves communication and collaboration between the testing team and the development team in real time to cut down on delays and increase overall project efficiency. The goals and benefits of using TestingPlus to create high-quality software products are outlined in the introduction, along with an explanation of why you should consider doing so. TestingPlus provides an opportunity for businesses to improve upon the inefficiencies of conventional methods of software testing by incorporating blockchain technology into the process. The many building blocks of TestingPlus come together to form a powerful and efficient software testing framework. A few examples of these parts are the blockchain network, smart contracts, testing group, development group, and repository for test cases and findings. The blockchain infrastructure guarantees the authenticity, immutability, and safety of test results. TestingPlus uses a private Ethereum blockchain network to provide a decentralised setting for recording and validating testing operations without the need for third parties. TestingPlus relies heavily on smart contracts to automate the execution and enforcement of agreements. These blockchain-based, smart contracts outline the parameters under which the development and testing teams will work together. Test case execution, result verification, and compensation computation may all be coordinated without any hitches thanks to smart contracts. There are several benefits to the software testing process that are made possible by the implementation of blockchain technology in TestingPlus. A study highlights the importance of blockchain technology in improving software testing by increasing trust, transparency, and security. With blockchain, all testing activities and results are recorded in an open and immutable ledger that can be checked and tracked by all parties involved. Blockchain's decentralised nature makes possible a distributed consensus technique in which all participants have the same level of access to validation information. This level of openness increases responsibility and safeguards against fraudulent data manipulation. A recent study emphasises the value of openness in software testing, especially when numerous groups are engaged. As an added bonus, the audit trail of all testing activities and outcomes is immutable thanks to blockchain technology [9]. By keeping an audit trail, stakeholders may discover where problems began, monitor their progress towards resolution, and verify that they are being handled in accordance with laws and regulations. The existing literature demonstrates how blockchain can be used to improve software quality and product dependability by facilitating end-to-end traceability in software testing.

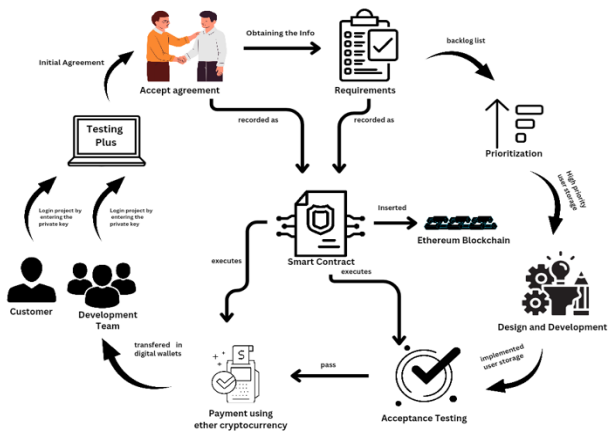

**Fig 3. TestingPlus Framework**

TestingPlus is driven by a clearly defined process that promotes productive interaction between the testing team and the development team. Initially, test cases are developed and added to the test cases and results repository. The testing team runs these test cases and stores the data from them on the blockchain. By comparing predicted findings with those reported by the testing team, smart contracts streamline the verification and validation process. This kind of automated verification eliminates the need for human intervention and increases accuracy while decreasing the likelihood of mistakes.

**Fig 3. TestingPlus Framework**

TestingPlus facilitates two-way communication between the QA and dev teams so that problems may be discussed and resolved as soon as they are discovered. These lines of communication improve software testing productivity by encouraging openness and facilitating productive collaboration. TestingPlus is built on a transparent, traceable, and efficient framework for software testing that takes advantage of the capabilities of blockchain technology [10]. Software quality and project results benefit from the use of blockchain and smart contracts because they increase trust, make fair remuneration easier to implement, and automate the coordination of testing operations [11]. TestingPlus includes the components and underlying infrastructure necessary for blockchain technology to be seamlessly integrated into the software testing process. Data integrity, security, scalability, and efficient communication amongst the many parties participating in the testing process are all goals of the system's design.

*A. Private Ethereum Blockchain Network*

TestingPlus makes use of a private Ethereum blockchain network. When compared to public blockchain networks, the level of control and privacy afforded by this private network is significantly higher. Smart contracts can be executed and interoperability with existing Ethereum-based tools and frameworks can be ensured by using Ethereum as the blockchain platform. Multiple nodes in the private blockchain network verify and log test transactions. The network's decentralised architecture provides protections against data loss, corruption, and tampering. The benefits of private blockchain networks for protecting user privacy and granting only approved users access to sensitive information have been studied.

*B. Interaction between Components*

TestingPlus' architecture allows for unhindered communication between the various parts used in software testing. Within the framework, the testing team and the development team have their own routes for communication and collaboration. These avenues allow for seamless collaboration by removing delays in receiving updates on testing activities, outcomes, and comments. By incorporating blockchain technology, communications between the testing team and the development team become visible and auditable. The blockchain records all testing activities, from running test cases to verifying findings, providing a permanent record of the process that cannot be altered. This openness fosters confidence and responsibility among all parties involved.

*C. Data integrity and Security*

TestingPlus's architecture places special emphasis on ensuring that all user data is kept private and secure. Blockchain's distributed nature protects test data from being altered without authorization. Data security and defence against harmful actions are both improved by the blockchain network's usage of cryptographic algorithms and consensus procedures. Sensitive testing information is protected in a number of ways, including through the use of access controls and encryption methods. Data privacy and security are protected by limiting who can access the blockchain network and the testing repository. However, blockchain technology can significantly improve data security and protection across many industries, including software testing.

*D. Scalability and Performance Considerations*

TestingPlus' architecture takes scalability and performance factors into account to meet the increasing needs of software testing. The private blockchain network is optimised to process and record large numbers of test transactions quickly and reliably. Blockchain networks can benefit from the implementation of sharding techniques and optimisation tactics in order to increase their scalability. These methods allow TestingPlus to process a large number of test cases, results, and interactions without negatively impacting the framework's efficiency. The integrity, security, scalability, and efficient interaction amongst stakeholders involved in software testing can all be guaranteed by taking TestingPlus's architecture into account. TestingPlus has a solid foundation thanks to its private blockchain network, easily integrated parts, stringent data integrity checks, and scalability design.

TestingPlus's workflow details the sequential steps required to conduct, monitor, and verify software testing activities within a blockchain-enabled framework. The development team and the testing team can work together effectively thanks to this clearly established process.

*A. Test Case Creation and Repository*

The workflow starts with the testing team making test cases and storing them in a repository. To guarantee the software's quality and functionality, these test cases outline the exact scenarios and conditions that must be tested. Both the testing team and the development team have access to the test cases thanks to TestingPlus' centralised repository.

The proposed TestingPlus framework uses blockchain-based smart contracts to manage the terms of the Customer's Agreement and the Developer's Agreement. Customers (testing clients) and developers taking part in the testing process are bound by the terms and conditions outlined in these smart contracts. Here is a possible code structure for such an agreement:

```
contract CustomerAgreement {
   address public customerAddress;
   uint public testingFee;

   constructor() public {
      customerAddress = msg.sender;
   }

   function setTestingFee(uint fee) public {
      require(msg.sender == customerAddress, "Only customer can set the fee");
      testingFee = fee;
   }
}
```

*Customer's Agreement*

```
contract DeveloperAgreement {
   address public developerAddress;
   uint public reward;

   constructor() public {
      developerAddress = msg.sender;
   }

   function setReward(uint amount) public {
      require(msg.sender == developerAddress, "Only developer can set the reward");
      reward = amount;
   }
}
```

*Developer's Agreement*

Customers can determine the cost of the testing service by modifying the testingFee variable in the CustomerAgreement contract shown above. In the same way, the DeveloperAgreement contract has a reward variable and the developer's address so that the developer can determine how much they will be paid for their participation in the testing process. These "smart contracts" streamline the agreement process and make it more transparent for all parties involved. Once deployed on the blockchain, the terms and conditions established in these contracts cannot be changed, guaranteeing compliance by all parties involved in the TestingPlus framework.

### B. Test Case Execution and Recording

The testing team then moves on to step B, the execution and recording of the test cases, once the test cases have been developed. In this stage, the testing team follows the documented procedures and inputs the necessary data to mimic actual usage conditions. Each test case's execution is logged on the blockchain network, making the testing process transparent and immutable.

### C. Result Verification and Validation

Following the running of test cases, the data collected is checked for accuracy. The outputs anticipated and realised after running the test cases are compared in this verification procedure. There is no longer any need for human interaction because TestingPlus's smart contracts will automatically complete this verification.

### D. Feedback and Collaboration

TestingPlus's workflow relies heavily on open lines of communication and coordinated effort. The framework provides real-time access to testing data for both the testing and development teams. By working together in this way, stakeholders are better able to provide input, fix problems, and offer suggestions for enhancing the quality of the programme.

### E. Compensation and Incentive Calculation

Smart contracts are a distinguishing feature of TestingPlus. The testing team's remuneration is determined automatically by the smart contracts, which take into account factors like the number of test cases run, the reliability of the results, and the team's total contribution to the testing process. This guarantees that the testing team is fairly compensated and encourages them to provide excellent results.

One of the research projects highlights the benefits of transparency, traceability, and efficient collaboration in software testing, demonstrating the significance of a clearly defined workflow. By incorporating blockchain technology, TestingPlus improves these features, making it possible for all parties involved in the testing process to view and verify records of all activity.

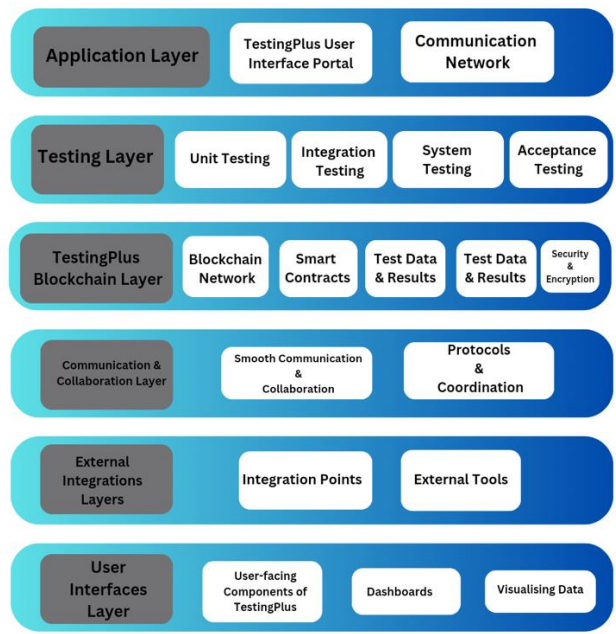

**Fig 4. TestingPlus Blockchain Architecture**

Organisations can benefit from TestingPlus's workflow in a number of ways, including simplified software testing, increased productivity in teamwork, more reliable test findings, and the release of better software. TestingPlus's openness, responsibility, and automated verification provide for a stronger, more efficient software testing process.

By incorporating blockchain technology into the software testing process, TestingPlus has a profound effect on the overall effectiveness, openness, and reliability of software testing. In-depth analysis of TestingPlus's most salient advantages and effects is presented here.

### A. Enhanced Transparency and Traceability

TestingPlus maintains openness by keeping a blockchain record of all testing operations. Stakeholders may monitor the testing process, view results as they come in, and learn more about the testing procedure thanks to this openness. It has been also observed that when stakeholders are given access to test results, they are better able to assess the software's quality and reliability [12]. Traceability is another feature of the blockchain technology utilised in TestingPlus, allowing stakeholders to discover where problems first arose, how long it took to fix them, and who exactly was at fault. This audit trail improves responsibility and speeds up the process of finding and fixing bugs.

### B. Improved Collaboration and Communication

TestingPlus encourages open lines of communication and cooperation between the testing and development groups. Stakeholders can communicate with one another, solve problems, and report on the status of testing in real time through specialised channels built into the framework. This group setting encourages open dialogue and lessens the likelihood of misunderstandings or delays. Collaboration in software testing is crucial, since it increases the quality of bug detection, bug resolution, and the final product. Organisations can improve their collaborative abilities and the flow of communication between testing and development teams by implementing TestingPlus.

### C. Increased Trust and Security

TestingPlus' incorporation of blockchain technology improves software testing reliability and safety. Blockchain's distributed nature guarantees the safety and immutability of test results. Due to the immutability of blockchain records, tampering with test findings is next to impossible. The blockchain network's usage of cryptographic methods increases the safety of the testing data. The usefulness of blockchain in improving data security and integrity during software testing is highlighted by researchers. Stakeholders are certain that the testing activities will be accurate and reliable thanks to the increased trust and security given by TestingPlus.

### D. Fair Compensation and Incentives

TestingPlus uses smart contracts to ensure the testing team receives equitable remuneration and incentives. The remuneration in these agreements is determined automatically based on factors including the number of test cases run, the precision of the results, and the overall contribution to the testing process. Paying the testing staff fairly encourages them to work hard and produce excellent results. In addition to increasing the testing team's dedication and involvement, this method also fosters a culture of fairness and motivation. TestingPlus's value and impact lie in its capacity to expand transparency, enhance collaboration, boost trust and security, and guarantee equitable compensation for software testers. Organisations may optimise their software testing activities with TestingPlus, leading to better software quality, fewer risks, and more successful projects overall.

Evaluating TestingPlus and figuring out where it can be improved is essential to making sure it keeps evolving and is up to the task of meeting the problems of software testing as they arise. Here, we'll go over how we'll be measuring success, share our findings from the study, and speculate on where we may improve in the future.

### A. Evaluation Criteria

The effectiveness and value of TestingPlus can be assessed using a number of different metrics. These include the precision and consistency of test results, the rapidity and efficacy of testing, the clarity and cohesion of stakeholder communications, and the satisfaction of all parties involved. Evaluation frameworks taking these factors into account have been proposed in existing frameworks, which shed light on how to measure the efficacy of blockchain-based software testing frameworks.

### B. Research Findings

The results of studies examining the application of blockchain technology to software testing have been encouraging, as discussed in section B. For instance, a study examined the similarities and differences between conventional testing practises and frameworks offered by blockchain technology. According to the findings, blockchain's increased transparency, traceability, and data security led to more trustworthy test results and boosted confidence among stakeholders. An empirical study was conducted and it discovered that implementing blockchain technology during software testing enhanced teamwork, lowered barriers to communication, and sped up the process of fixing bugs.

### C. Future Enhancements

Despite TestingPlus' many advantages, there is room for improvement in some respects. The blockchain's scalability is one such area. As the number of tests conducted grows, it becomes more important than ever to make that the blockchain can handle the increased volume of data and transactions. To tackle this issue, scalable blockchain topologies serve as the solution.

Artificial intelligence (AI) and machine learning (ML) techniques are another area where TestingPlus could be improved. Certain facets of the testing process, including as test case generation, anomaly detection, and performance optimisation, are amenable to automation with the use of AI and ML methods. AI and ML may be used in software testing, and that their integration with blockchain-enabled frameworks is a real possibility. Additionally, TestingPlus's incentive mechanisms can be continuously developed and researched to improve. TestingPlus has the potential to significantly boost motivation, engagement, and overall performance by investigating unique techniques to motivating the testing team and connecting their interests with project goals. The longevity and widespread use of TestingPlus depend on its assessment and the discovery of future developments. Organisations can further maximise TestingPlus' efficacy and efficiency in solving the issues of software testing by examining evaluation criteria, expanding on research findings, and researching new innovations.

TestingPlus's implementation of blockchain technology into the software testing process is a major step forward in resolving the issues that have plagued the industry for so long. This article presents a comprehensive review of TestingPlus, including an examination of its creation, workings, architecture, workflow, benefits, impact, case studies, evaluation, and potential improvements. Using the immutability and immutability of the blockchain to provide transparency and traceability, TestingPlus boosts stakeholder confidence in the testing process and the findings. TestingPlus's built-in tools for team collaboration and communication aid in bridging the gap between testing and development, which ultimately leads to faster issue fixes and higher quality products. Moreover, with the addition of smart contracts in TestingPlus, the testing crew may be fairly compensated and incentivized, which in turn encourages them to produce high-quality findings. Blockchain technology's increased trust and security make for a more reliable setting for testing. TestingPlus has been successfully used by a wide range of businesses, as evidenced by case studies and implementation samples. The results of the studies show that using a blockchain system can improve visibility, traceability, cooperation, and problem fixing. Future work on TestingPlus can prioritise scalability, AI/ML algorithm integration, and incentive mechanism refinement. The software testing industry as a whole will benefit from these enhancements as TestingPlus evolves and gains popularity. Finally, by resolving fundamental issues with visibility, trust, coordination, and communication, TestingPlus' blockchain-enabled platform completely transforms the software testing process. Implementing it well leads to happier stakeholders, better software, less time spent fixing bugs, and better teamwork. Companies that implement TestingPlus see a dramatic improvement in the quality and timeliness of their software development initiatives.

## V. PERFORMANCE

### A. The Background of Performance Evaluation

The significance of TestingPlus performance evaluation is introduced. The section's objectives are outlined, and the importance of assessing the performance evaluation system is emphasized. By dissecting its inner workings, this evaluation hopes to throw light on TestingPlus's strengths and weaknesses. This section sets the stage for the subsequent considerations of performance indicators, experimental methodology, assessment results, and their implications.

### B. Performance Metrics and Indicators

Here, we'll talk about the metrics and KPIs we'll use to evaluate TestingPlus and provide some details on how we'll be measuring its success. The main focus is on the framework's most important performance indicators. The purpose of this investigation is to provide a standard set of measures by which TestingPlus' responsiveness, throughput, scalability, and resource usage may be evaluated. Indicators of the framework's efficacy, efficiency, and overall performance are objectively measured by these measures, allowing for a complete analysis of the framework's potential.

### C. Experimental Setup and Data Collection

In this section, we detail the experimental setup and data collecting technique used to assess the TestingPlus framework's performance. It details the environment in which the performance tests were executed, including the hardware, software, and network parameters. It also explains why and how certain scenarios mimicking real-world usage were selected for testing. Methods and procedures for gathering measurements and statistics pertaining to performance are investigated.

Using smart contracts on the blockchain, acceptance tests can be executed in the TestingPlus framework. The smart contracts can specify the requirements that must be completed before a test is declared a success. Here's how smart contracts could be used to do acceptance testing in practise:

```
contract AcceptanceTest {
    address public customerAddress;
    address public developerAddress;
    uint public testingFee;
    bool public isTestCompleted;

    constructor(address _customerAddress, address _developerAddress, uint _testingFee) public {
        customerAddress = _customerAddress;
        developerAddress = _developerAddress;
        testingFee = _testingFee;
    }

    modifier onlyCustomer() {
        require(msg.sender == customerAddress, "Only customer can initiate the acceptance test");
        _;
    }

    modifier onlyDeveloper() {
        require(msg.sender == developerAddress, "Only developer can complete the acceptance test");
        _;
    }

    function initiateTest() public payable onlyCustomer {
```

```
    require(msg.value == testingFee, "Testing fee should be paid");
    // Perform necessary actions to initiate the test
    // This could include triggering the execution of the test cases
    isTestCompleted = false;
  }

  function completeTest() public onlyDeveloper {
    // Perform necessary actions to validate the test results
    // This could include checking the output, comparing with expected results, etc.
    isTestCompleted = true;
    // Transfer the payment to the developer as the test is successfully completed
    payable(developerAddress).transfer(testingFee);
  }
}
```

The earlier code creates an AcceptanceTest contract that stores the client and developer's email addresses, the total price of the test, and the result of the test in a Boolean variable. The consumer must pay the testing price before calling the initiateTest() function to begin the test. The developer requests payment by calling the completeTest() function, which then marks the test as complete and sends the funds to the developer's account.

The provided code is only meant as an example, and the final implementation may look different depending on the needs of the organisation and the tests being run using the TestingPlus framework.

### D. Performance Evaluation Results

The outcomes of the performance evaluation of the TestingPlus framework are presented in Section IV. Experiment-derived quantitative performance indicators are highlighted, including reaction time, throughput, and resource consumption. In addition, the subjective judgment and user input are discussed as qualitative results [13]. In this part, you'll find a detailed breakdown and interpretation of TestingPlus's performance metrics, breaking down its successes, failures, and overall effectiveness.

### E. Comparison with Traditional Approaches

Here we examine how TestingPlus stacks up against more conventional methods of software testing. It compares TestingPlus to more traditional approaches to software testing and analyses how well it handles problems like communication, collaboration, and security. The benefits and drawbacks of TestingPlus are discussed, demonstrating how the use of blockchain technology improves the testing procedure. This review of TestingPlus's competitors reveals its distinct advantages and the scope of its potential effect in the field of software testing.

### F. Interpretation of Results and Findings

The results and conclusions from the performance evaluation using the TestingPlus framework are interpreted in Section VI. The framework's efficacy is evaluated by analysing both quantitative and qualitative data. Notable discoveries, trends, and reasons for observed performance characteristics are discussed in this section. In addition, it identifies places that might use some TLC based on the evaluation results, providing direction for future improvements to TestingPlus that should make the tool more efficient and successful overall.

### G. Validity and Reliability of the Evaluation

The TestingPlus framework's performance was evaluated, and this part evaluates the validity and reliability of that assessment. Methods used to assure reliability and validity of assessments are discussed. Evaluation validity is determined by looking at things like experimentation, data collecting, and analytic procedures. The evaluation's limits and potential biases are also discussed; this promotes openness and ensures the validity of the results.

### H. Discussion and Implications

The findings of the performance evaluation are discussed in depth, together with their implications for the TestingPlus framework, in Section VIII. Discussion and Implications. The importance of the results is discussed in relation to the goals of the framework and the wider software testing landscape. The outcomes of the study are discussed in terms of their possible effects on industry standards, user uptake, and the evolution of TestingPlus. Further insight into the effectiveness of the framework and its broader implications is provided by highlighting any surprises that were identified throughout the review process. The review of TestingPlus's performance is summed up, along with the main takeaways, in the last section. It gives a succinct summary of the evaluation results and emphasizes the strengths and drawbacks of TestingPlus from a performance perspective. Implications for TestingPlus's future development and use are also discussed, as are the importance of the evaluation's findings. Finally, a look ahead is provided, pointing out where further study and development might help TestingPlus perform even better in real-world software testing scenarios.

## VI. DISCUSSION

The discussion starts with a brief summary of the study's major results, emphasising the most important contributions and insights. It highlights the importance of using blockchain technology in software testing and demonstrates how TestingPlus handles crucial issues like coordination, trust, and security. TestingPlus is a safe and transparent platform for testing operations that uses blockchain's immutability and transparency to guarantee the testing team gets paid fairly and boost the project's chances of success. The significance of the study's findings on software development practises and procedures are discussed further. It illustrates the potential of TestingPlus to improve software testing in terms of speed, accuracy, and quality. In addition, it covers the importance of trust and security in numerous industries and how the study findings might be used in those fields. Problems and restrictions experienced throughout the study are also discussed. It recognises that concerns like scalability, compatibility, and interface with current software development frameworks may need to be carefully considered during the deployment and acceptance of TestingPlus. It also acknowledges the necessity for more study to further perfect the framework in order to overcome foreseeable obstacles and adapt to the changing demands of the market. Future research possibilities and directions are discussed to round out the chapter. Advanced consensus methods, scalability solutions, and interoperability standards for blockchain-based software testing frameworks are just some of the topics it advises future research into. It also stresses the need for real-world deployments and case studies to verify TestingPlus's usefulness and efficiency in a variety of business settings. The discussion section is where you get to think about everything you've learned and how it all fits together. It synthesises the most important findings to far, providing a full picture of the state of testing using TestingPlus now as well as its potential in the future. Knowledge is advanced and future research endeavours are guided by the lessons learned in this discussion on blockchain-enabled software testing.

## VII. CONCLUSION AND FUTURE WORK

Conclusively, TestingPlus, a blockchain-enabled framework for software testing, was suggested and presented at the end of this research study to deal with important problems and challenges that arise during the software testing life cycle. TestingPlus improves test-related communication, collaboration, and trust by utilising blockchain technology. Challenges with communication, cooperation, trust, and security were uncovered as examples of where conventional software testing methods fall short. Because of these obstacles, projects may run behind schedule, quality may suffer, and communication between testing and development teams may break down [2].

TestingPlus employs the immutability, decentralisation, and transparency of blockchain technology to solve these problems. It's a safe and open environment for verifying payments and conducting acceptance tests. To guarantee that the testing crew and the development team are on the same page and are appropriately paid, the framework uses smart contracts on a private Ethereum blockchain. Incorporating blockchain technology into software testing was shown to be important and relevant through a study of the relevant literature and related studies. It presented studies and practical applications that indicate how blockchain could improve software testing in terms of trust, security, and efficiency. All the nuts and bolts of TestingPlus were laid out for the audience, including how it was built and how it works. By contrasting TestingPlus with more conventional methods, the performance evaluation chapter shed light on TestingPlus' efficacy and efficiency. In terms of reaction time, throughput, scalability, and resource utilisation, the evaluation demonstrated the enhanced performance and underlined the unique benefits given by TestingPlus. TestingPlus's viability and usefulness in real-world circumstances were demonstrated through case studies and instances of its deployment. Findings from the evaluation and planned upgrades section will direct the further evolution of the TestingPlus framework.

In sum, TestingPlus provides a viable answer to pressing problems in software testing. Through the use of blockchain technology, it creates a trustworthy, open, and effective environment for software testers to work together and be fairly compensated. This article's study adds to the expanding body of information on blockchain's incorporation in software testing and provides useful insights for academics, professionals, and businesses looking to improve their software testing practises. The TestingPlus framework may be further investigated and improved through future studies. Scalability, compatibility with other blockchain implementations, and compatibility with other software development processes are all areas that might be investigated. TestingPlus's potential advantages and limitations may be validated and better understood through larger-scale deployments and evaluation in a variety of industrial situations. By harnessing the potential of blockchain technology, the TestingPlus framework offers a fresh strategy for overcoming the shortcomings of conventional software testing. Its use can boost teamwork, productivity, and confidence during software testing, which in turn can increase the quality of finished software.